\RequirePackage{lineno}
\documentclass[prd,twocolumn,showpacs,preprintnumbers,floatfix,amsmath,amssymb,amsfonts]{revtex4}

\usepackage{graphicx}
\usepackage{amsmath}
\usepackage{dcolumn}
\usepackage{epsfig}
\usepackage{psfrag}
\usepackage{subfigure}
\usepackage{hyperref}

\RequirePackage{xspace}

\def\TeV{\ifmmode {\mathrm{\ Te\kern -0.1em V}}\else
	                   \textrm{Te\kern -0.1em V}\fi}%
\def\GeV{\ifmmode {\mathrm{\ Ge\kern -0.1em V}}\else
	                   \textrm{Ge\kern -0.1em V}\fi}%
\def\MeV{\ifmmode {\mathrm{\ Me\kern -0.1em V}}\else
	                   \textrm{Me\kern -0.1em V}\fi}%
\def\keV{\ifmmode {\mathrm{\ ke\kern -0.1em V}}\else
	                   \textrm{ke\kern -0.1em V}\fi}%
\def\eV{\ifmmode  {\mathrm{\ e\kern -0.1em V}}\else
	                   \textrm{e\kern -0.1em V}\fi}%
\let\tev=\TeV
\let\gev=\GeV

\def\TeVc{\ifmmode {\mathrm{\ Te\kern -0.1em V}/c}\else
	                   {\textrm{Te\kern -0.1em V}/$c$}\fi}%
\def\GeVc{\ifmmode {\mathrm{\ Ge\kern -0.1em V}/c}\else
	                   {\textrm{Ge\kern -0.1em V}/$c$}\fi}%
\def\MeVc{\ifmmode {\mathrm{\ Me\kern -0.1em V}/c}\else
	                   {\textrm{Me\kern -0.1em V}/$c$}\fi}%
\def\keVc{\ifmmode {\mathrm{\ ke\kern -0.1em V}/c}\else
	                   {\textrm{ke\kern -0.1em V}/$c$}\fi}%
\def\eVc{\ifmmode  {\mathrm{\ e\kern -0.1em V}/c}\else
	                   {\textrm{e\kern -0.1em V}/$c$}\fi}%
	
\def\cm{\ifmmode  {\mathrm{\ cm}}\else	                   
\textrm{~cm}\fi}%

\def\mm{\ifmmode  {\mathrm{\ mm}}\else	                   
\textrm{~mm}\fi}%
	
\def\ifb{\mbox{fb$^{-1}$}}

\usepackage{relsize}
\def\babar{\mbox{\slshape B\kern-0.1em{\smaller A}\kern-0.1em
    B\kern-0.1em{\smaller A\kern-0.2em R}}}

\begin{document}

\title{Identifying boosted hadronically decaying top quark using jet \\substructure in its  center-of-mass frame}

\author{Chunhui Chen} 
\affiliation{Department of Physics and Astronomy, Iowa State University, Ames, Iowa 50011, USA}

\begin{abstract}
In this paper we study the identification of boosted hadronically decaying top quarks
using jet substructure in the center-of-mass frame of the jet. 
We demonstrate that the method can greatly reduce the QCD jet background while maintaining high
identification efficiency of the boosted top quark even in a very high
pileup condition. Applications to searches for heavy resonances that decay to a $t\bar{t}$ final state are also discussed.
\end{abstract}

\maketitle

\section{Introduction}
\label{sec:intro}
Search for and discovering new physics (NP) beyond the standard model (SM) has been
one the of main physics motivations to build the Large Hadron Collider (LHC).
One prominent path to find NP is through model independent  searches for possible
new particles beyond the SM. Many extensions of the SM predict new heavy resonances with 
masses at the TeV scale.  Some of these heavy resonances, such  as a  new heavy gauge boson 
$Z^\prime$ or Kaluza-Klein (KK) gluons from the bulk Randall-Sundrum model, can predominantly decay to a top-anti-top quark 
pair ($t\bar{t}$) final state~\cite{Agashe:2006hk,Contino:2006nn,Matsumoto:2006ws,Fitzpatrick:2007qr}. 
Because of the energy scale of these processes, the top quarks from the heavy resonance decay 
receive a significant Lorentz boost. Their hadronically decaying  products are usually so 
collimated that they can be only reconstructed as single jets in the experiments.

In recent years, many theoretical  studies have been performed to investigate the signature 
of boosted hadronically decaying top quarks~\cite{Thaler:2008ju,Kaplan:2008ie,Almeida:2008tp,Krohn:2009wm,Plehn:2010st,Chekanov:2010vc,Bhattacherjee:2010za,Rehermann:2010vq,Chekanov:2010gv,Jankowiak:2011qa,Thaler:2011gf,Soper:2012pb}, where a hadronically decaying top is defined as the top quark whose $W$ boson daughter 
decays hadronically.  Several experimental searches for heavy new resonances 
decaying to $t\bar{t}$ final states have  also been carried out by the ATLAS and CMS experiments at the LHC~\cite{Aad:2012raa,Chatrchyan:2012ku}.
The measurements exclude a production of such a new heavy resonance with a mass up to 1-2\,\tev, 
depending on the NP model used for the theoretical interoperation of the results.
In all those studies and measurements,  the complete final state of the top quark decay is reconstructed as a single jet, hereafter referred as $t$ jet.  
The invariant mass
of the reconstructed jet ($m_{\rm jet}$) is used to distinguish the $t$ jets from 
 QCD jets, where the QCD jets are defined as those jets initiated by a non-top quark or gluon.
 Since the jet mass alone may not provide sufficient discriminating power to
effectively separate $t$ jets from the overwhelming QCD background in many analyses,
techniques based on jet substructure information, such as jet shape observables~\cite{Abdesselam:2010pt},
filtering~\cite{Butterworth:2008iy}, pruning~\cite{Ellis:2009su,Ellis:2009me} and trimming~\cite{Krohn:2009th}, 
are typically implemented as additional experimental handles to  identify boosted hadronically decaying $t$ quarks.

In our previous paper~\cite{Chen:2011ah}, we introduce a new approach to study jet substructure 
in the center-of-mass frame of the jet. We demonstrated that it can be used to discriminate the boosted heavy particles
from the QCD jets and the method is complementary to other jet substructure algorithms.
A similar idea has also been explored to search for hadronically decaying Higgs boson~\cite{Kim:2010uj}.
In this paper, we will extend the studies presented in Ref.~\cite{Chen:2011ah} to focus on identifying 
the boosted hadronically decaying top quark in the center-of-mass frame of the jet. 
We demonstrate that the method can greatly reduce the QCD jet background while maintaining a high
identification efficiency of the boosted top quark even in the environment of a very large number 
of multiple interactions per event (pileup).
Using an example of applications, we show a good prospective on search for heavy mass particles in the 
$t\bar{t}$ decay channel at the LHC.

This paper is organized as follows: In Section~\ref{sec:sample}, we describe the event sample used  in the study.
Section~\ref{sec:jet_sub} discusses the method to identify $t$ jets using  jet substructure in the jet center-of-mass frame and its performance.
An example of the application of our method is given in Section~\ref{sec:app}.
We conclude in Section~\ref{sec:conclusion}.

\section{Event Sample}
\label{sec:sample}
We use boosted $t$ jets, from the SM process of 
$t\bar{t}$ production,  as a benchmark to study the identification of
boosted hadronically decaying top quark with our proposed jet substructure method. 
For simplicity we only consider the background from the SM dijet production since its cross section 
is several orders of magnitudes larger than those of other SM background. In addition, we also generate 
events to simulate a heavy-particle $X$ that decays to a $t\bar{t}$ final state.

All the events used in this analysis are produced using the P{\footnotesize ythia} 6.421 event generator~\cite{Sjostrand:2006za}
for the $pp$ collision at $14\,\rm TeV$ center-of-mass energy.
In order to simulate the finite resolution of the
Calorimeter detector at the LHC, we divide the $(\eta, \phi)$ plane into $0.1\times 0.1$ cells. We sum over the energy
of particles entering each cell in each event, except for the neutrinos and muons, and replace it with a massless pseudoparticle
of the same energy, also referred to an energy cluster,  pointing to the center of the  cell. These pseudoparticles are fed into the 
F{\footnotesize astJet} 3.0.1~\cite{fastjet} package for  jet reconstruction.
The jets are reconstructed using the anti-$k_T$ algorithm~\cite{Cacciari:2008gp}  
with a distance parameter of $\Delta R=0.6$. The  anti-$k_T$ jet algorithm is the default one used at the ATLAS and CMS experiments. 
In order to evaluate the performance of top jet identification with the currently expected experimental conditions at the LHC,
we generate MC events with different average numbers of multiple interactions per event~\cite{Sjostrand:2006za}
and
then repeat our studies for each scenarios. We 
compare the results to the one in the ideal experimental condition that has no pileup.

\section{Jet Substructure in the rest frame}
\label{sec:jet_sub}
In this section we describe the method to study jet substructure of the $t$ jets in the center-of-mass frame of
the jet in order to distinguish them  from the QCD jets. 
We select jets with $p_{\rm T}\ge600\,\gev$ and $|\eta|\le1.9$ as $t$ jet candidates,
where $p_{\rm T}$ and $\eta$ are the transverse momentum and pseudorapidity of the jet.
We further require that the $t$ jet candidates have $50\,\gev\le m_{\rm jet}\le 350\,\gev$. 
In case there are more than one candidate in an event, all of them are kept for further analysis.

\subsection{Center-of-mass  frame of  a jet}
We define the center-of-mass frame (rest frame) of a jet as the frame where the four momentum of the
jet is equal to $p^{\rm rest}_{\mu}\equiv (m_{\rm jet}, 0, 0, 0)$. A jet consists of its constituent particles.
The distribution of the constituent particles of a boosted $t$ jet in its center-of-mass frame
has a three body decay topology as in the top quark rest frame. On the other hand, 
the constituent particle distribution of a QCD jet in its rest frame does not correspond 
to any physical state and is more likely to be isotropic. 

\subsection{Reclustering in the jet reset frame}
We recluster the energy clusters
of a jet to reconstruct subjets in the jet rest frame. The reclustering is done by using the 
Cambridge-Aachen (CA) sequential jet reconstruction algorithm~\cite{Dokshitzer:1997in}  with a modified distance parameter 
of $\Delta\theta = 0.6$, where $\theta$ is defined as the angle between two pseudoparticles in the jet rest frame.
We reject jets that have less than 3 subjets with $E_{\rm jet}>10\,\gev$, where $E_{\rm jet}$ is
the energy of a subjet in the jet reset frame. In the ideal situation without any pileup effects, this
requirement rejects roughly 60\,\% of the QCD jets, while keeping almost all the signal $t$ jets.
However, the rejection power drops significantly when the average numbers of multiple interactions per event increases.
When the average pileup at the LHC reaches 50 (100) per event, more than 70\,\% (90\,\%) of the QCD jets
have at least 3 subjets with $E_{\rm jet}>10\,\gev$. 

\begin{figure}[!htb]
\begin{center}
\includegraphics[width=0.22\textwidth]{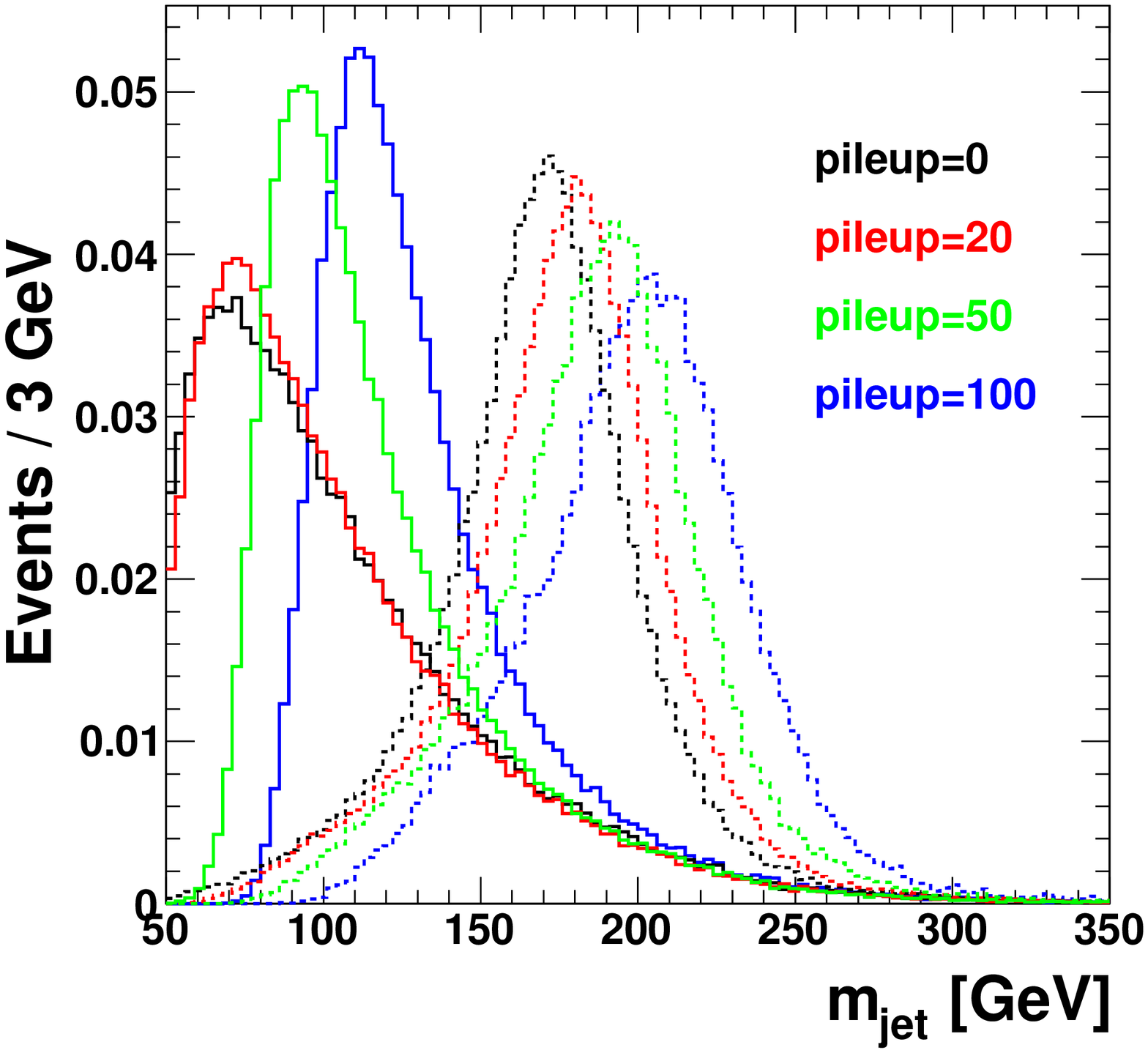}
\includegraphics[width=0.22\textwidth]{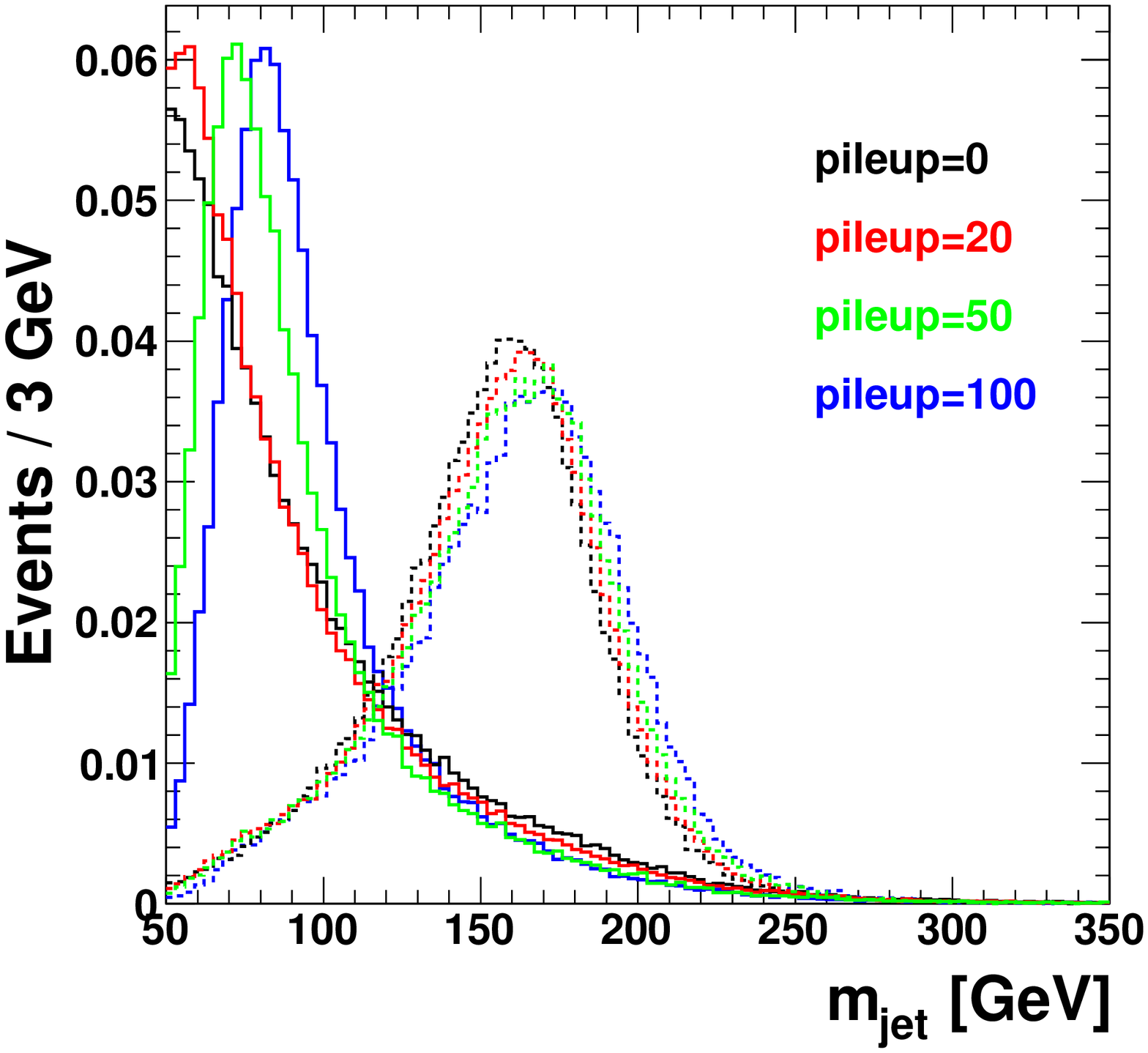}
\caption{The jet mass distributions of the QCD jets (solid line) and signal $t$ jets (dashed line) of the SM $t\bar{t}$ production 
from MC simulated events before (left) and after (right) reclustering in the jet rest frame under different pileup conditions. 
The mass of the jet after the reclustering is calculated using the 
three subjets with the highest energies in the jet rest frame. All the distributions are normalized to
unity.}
\label{fig:mjet}
\end{center}
\end{figure}

The mass of a jet has a large
dependence on the pileup condition, its distribution shifts to higher values when the pileup 
condition gets worse. However, this effect is greatly reduced if we calculate the mass of a jet
using its subjets in the center of mass frame of the jet, as shown in Fig.~\ref{fig:mjet}. 
Because of the smaller area covered by its cone size and the isotropic distribution of the 
constituent particles of QCD jets from pileup  in the jet rest frame,
a reconstructed subjet in the jet rest frame includes
much less deposited energies from additional multiple interactions.
One advantage of reclustering in the jet rest frame is  that two nearby energy clusters 
that have different momenta in the lab frame can be easily separated geometrically after they are boosted back
to the center-of-mass frame of the jet if they are originated by different partons from top decaying. 
We also point out that the rest frame subjet algorithm is infrared and collinear safe if an infrared
and collinear safe jet algorithm is used for the rest frame subjet clustering. All the sophisticated jet-grooming 
algorithms introduced in the lab frame, such as pruning~\cite{Ellis:2009su,Ellis:2009me} and trimming~\cite{Krohn:2009th},
can be easily incorporated. 

\begin{figure*}[p]
\begin{center}
\includegraphics[width=0.24\textwidth]{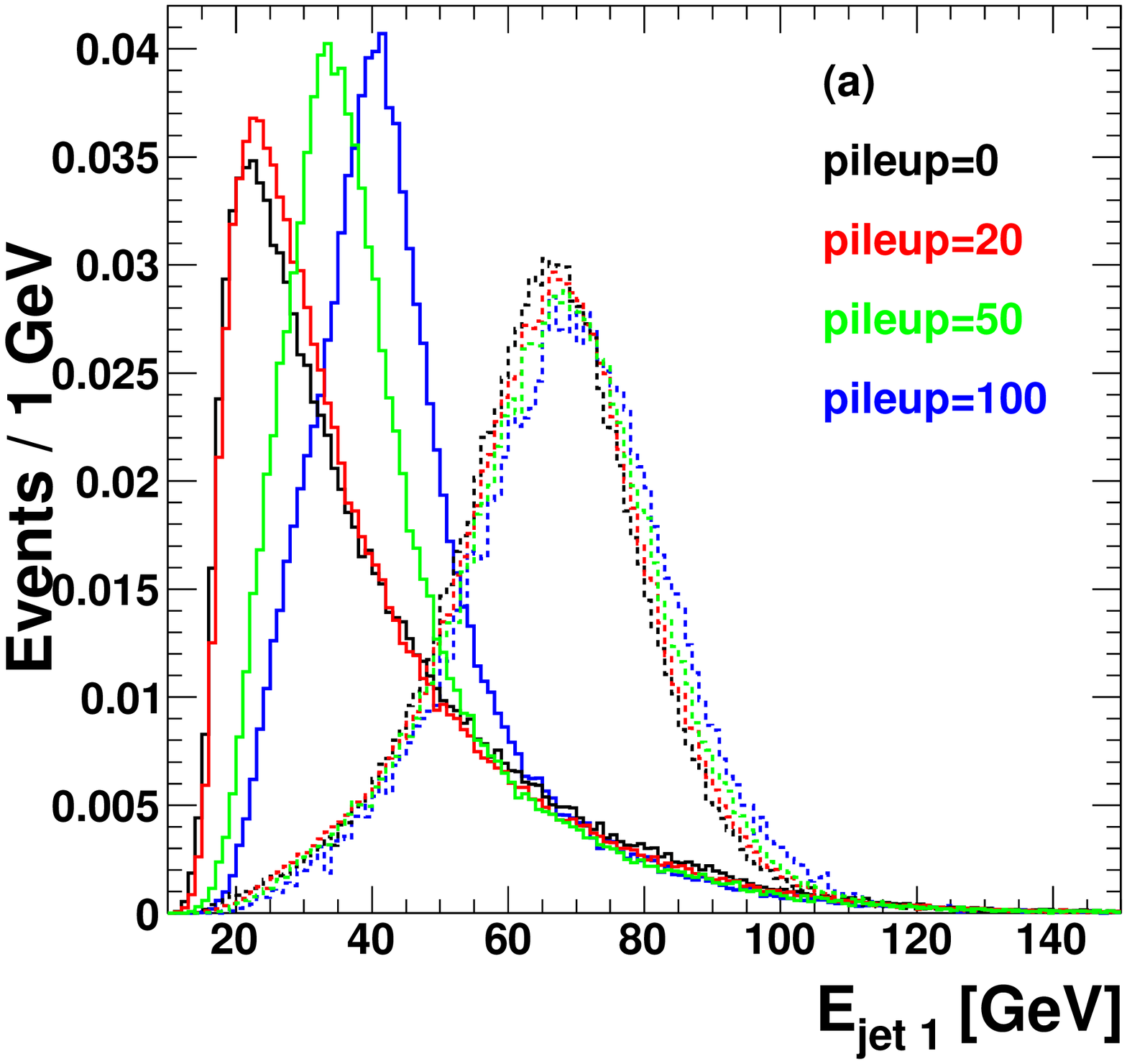}
\includegraphics[width=0.24\textwidth]{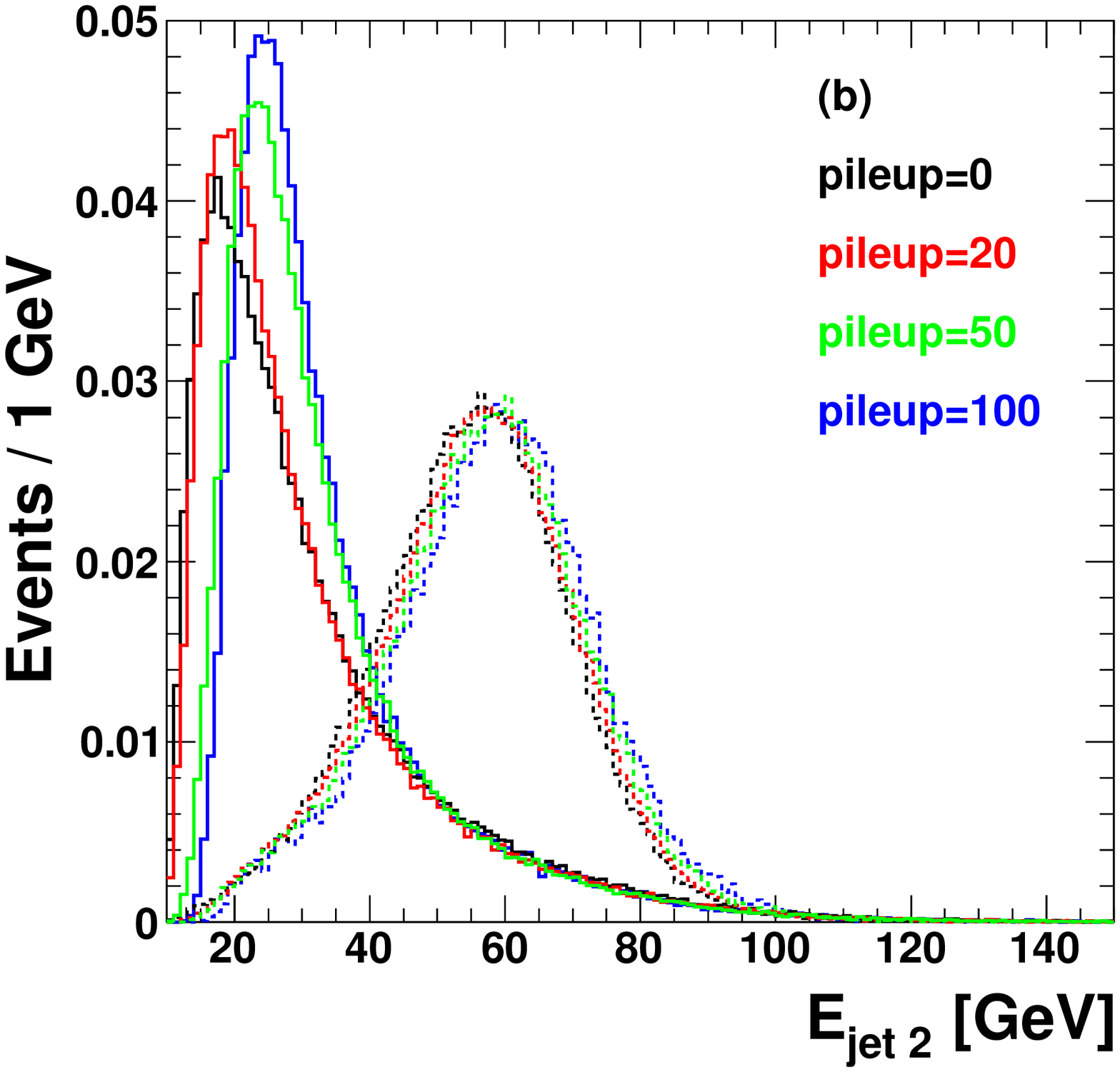}
\includegraphics[width=0.24\textwidth]{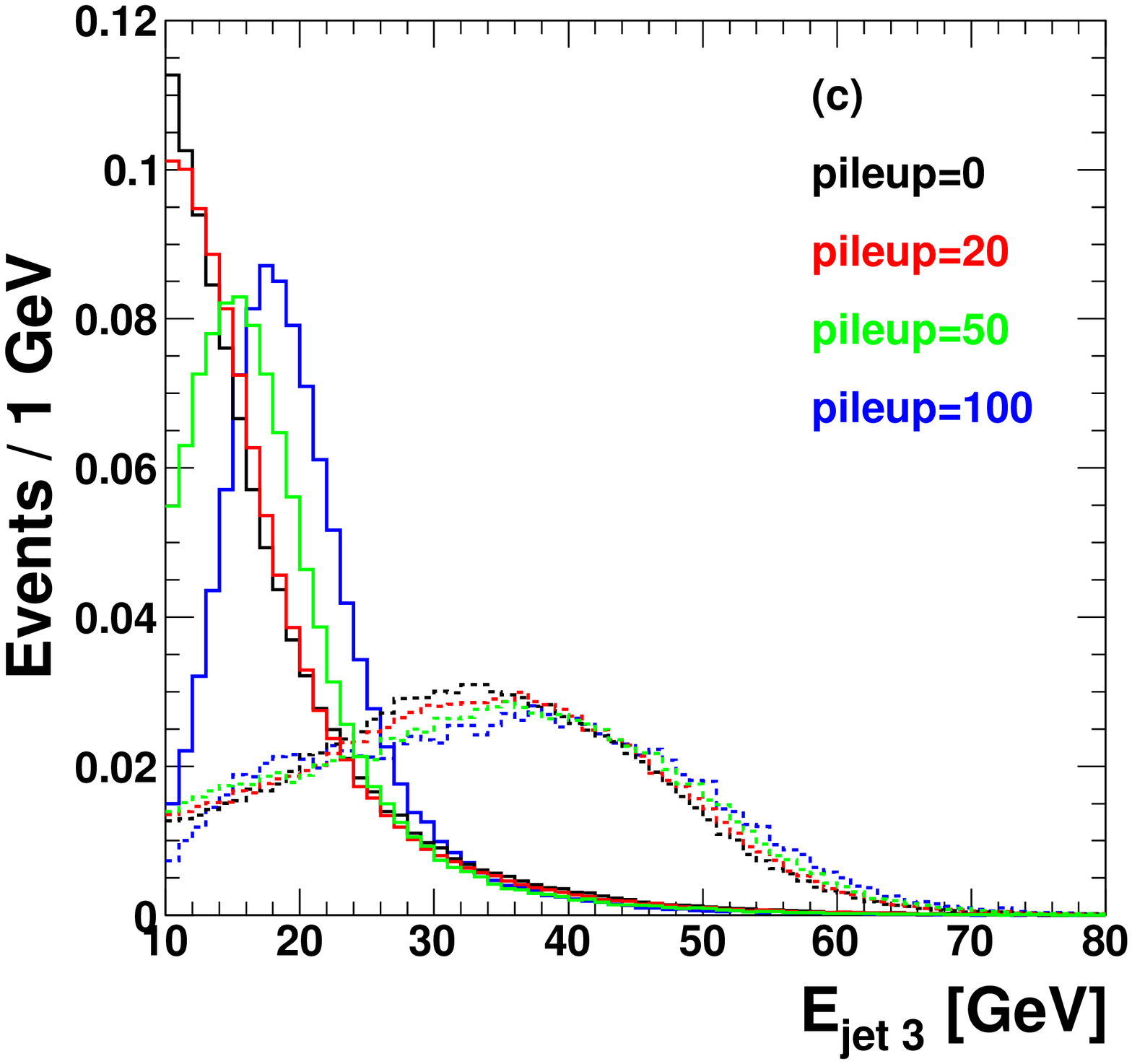}
\includegraphics[width=0.24\textwidth]{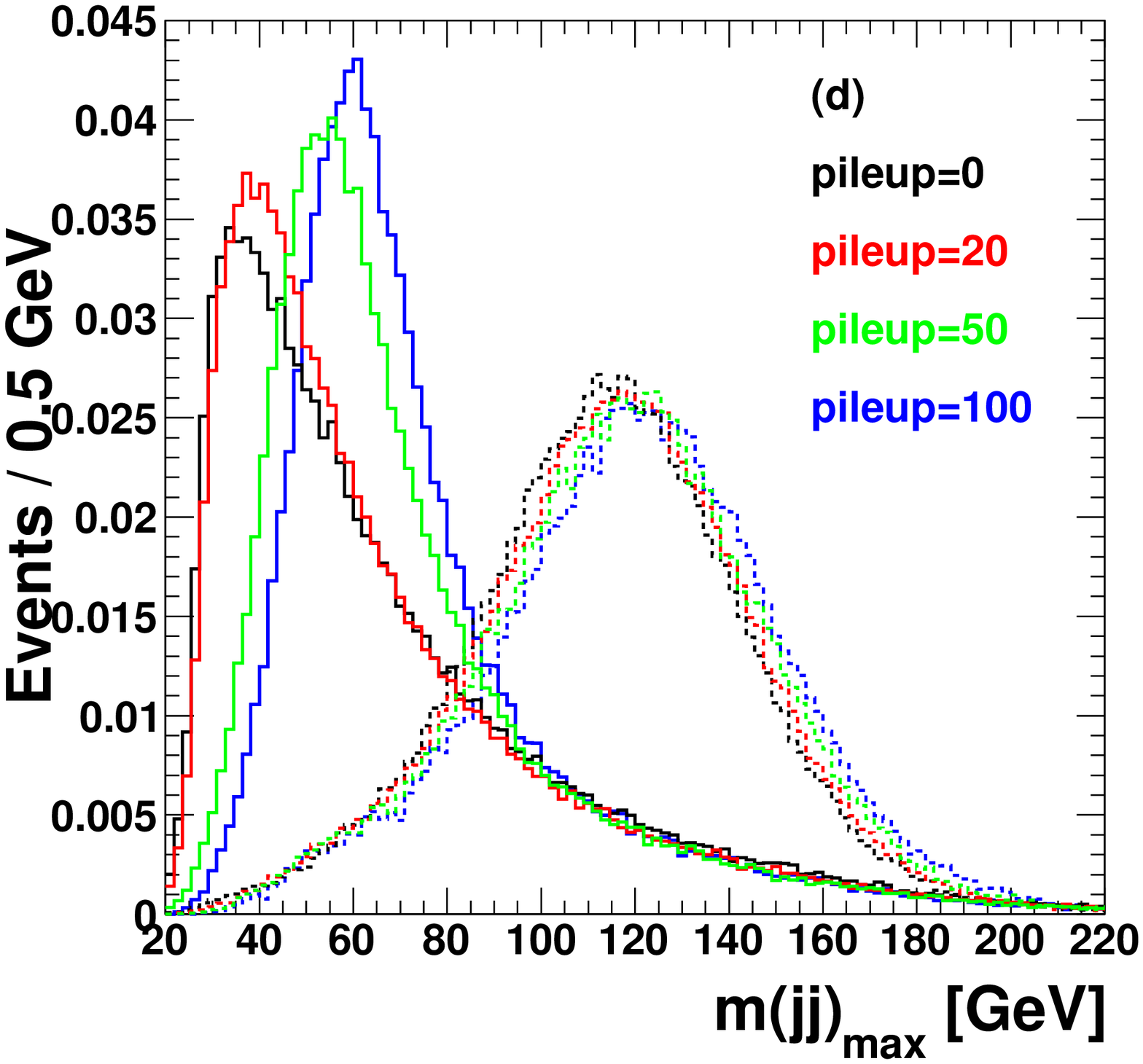}
\includegraphics[width=0.24\textwidth]{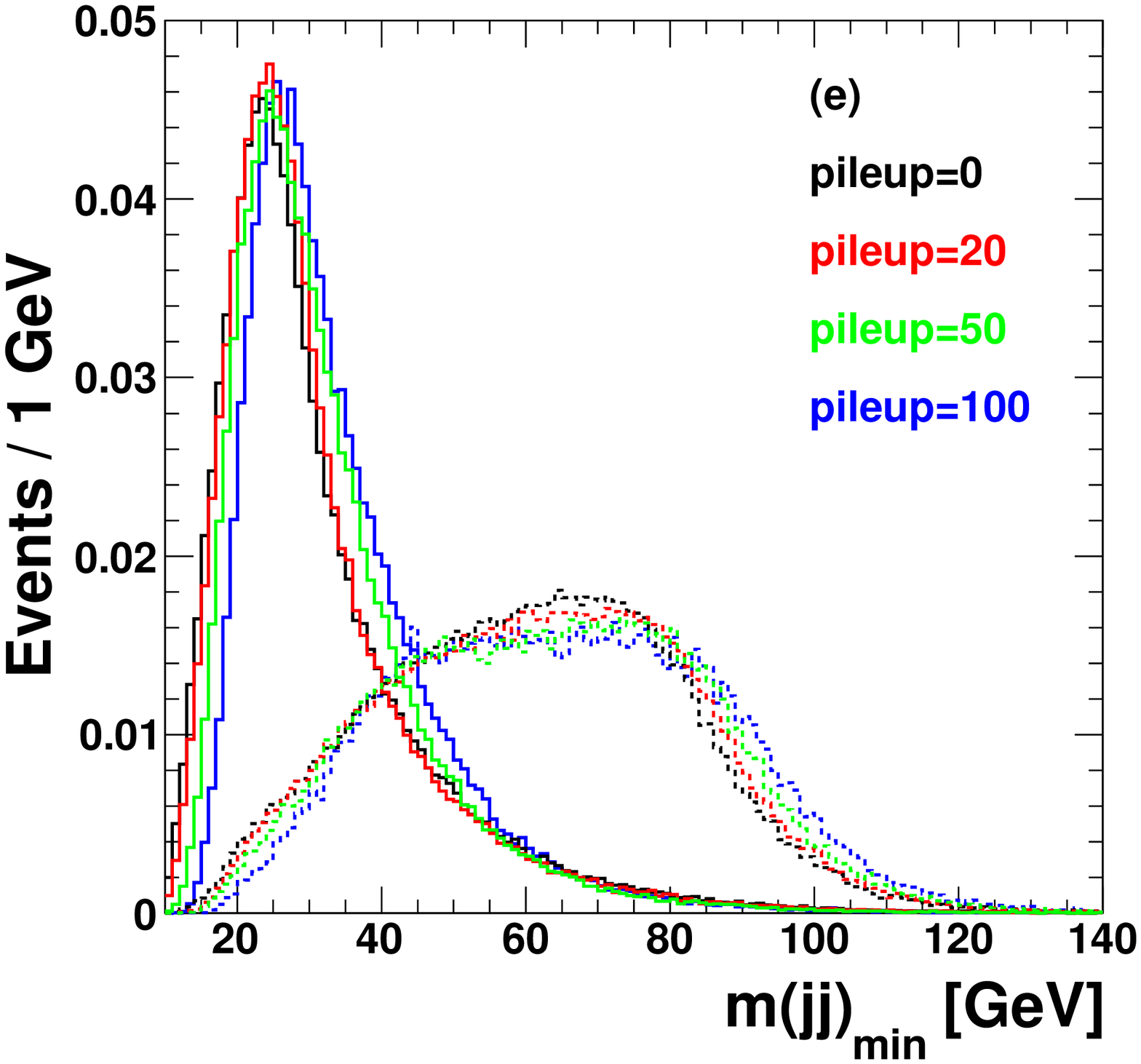}
\includegraphics[width=0.24\textwidth]{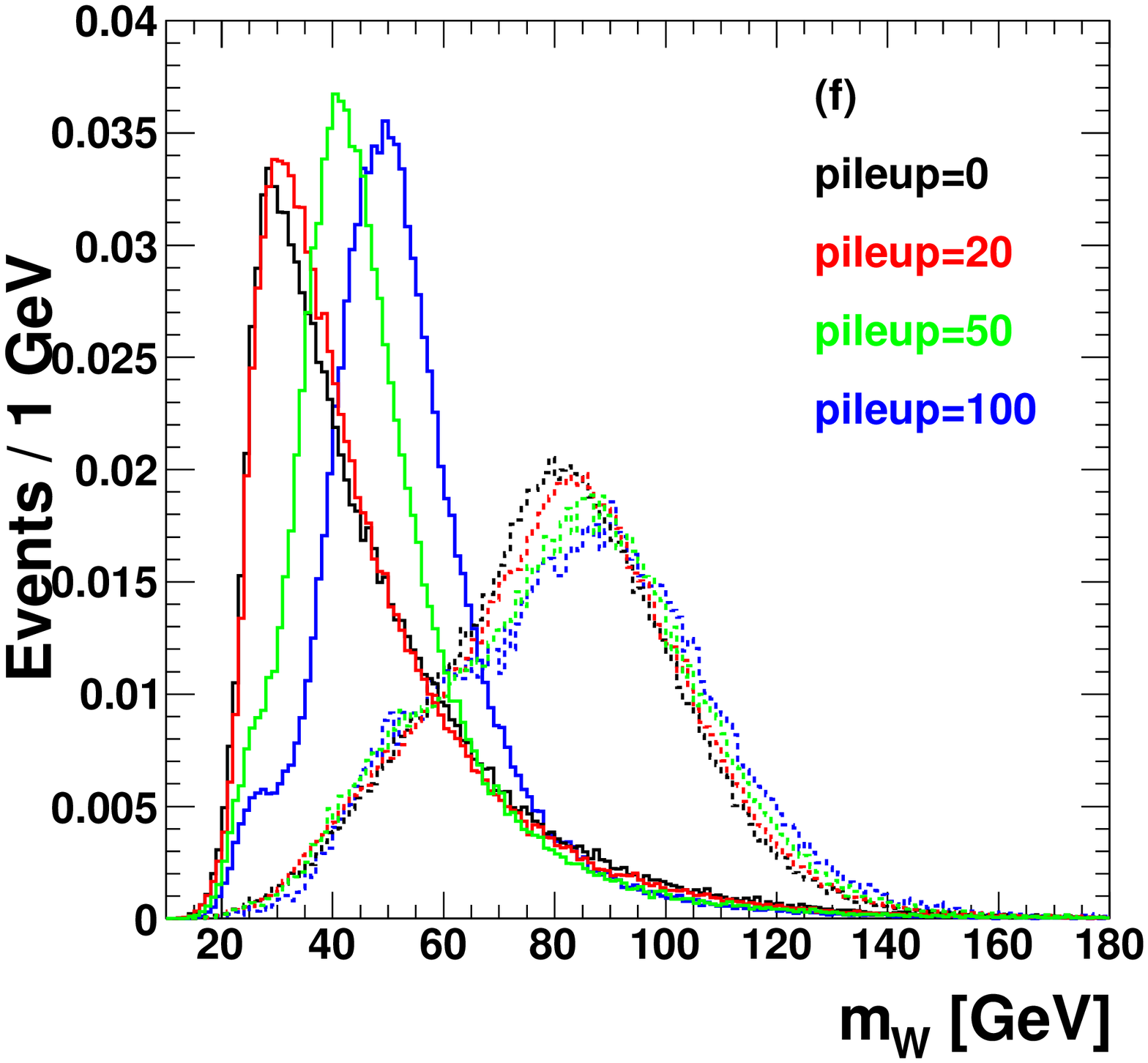}
\includegraphics[width=0.24\textwidth]{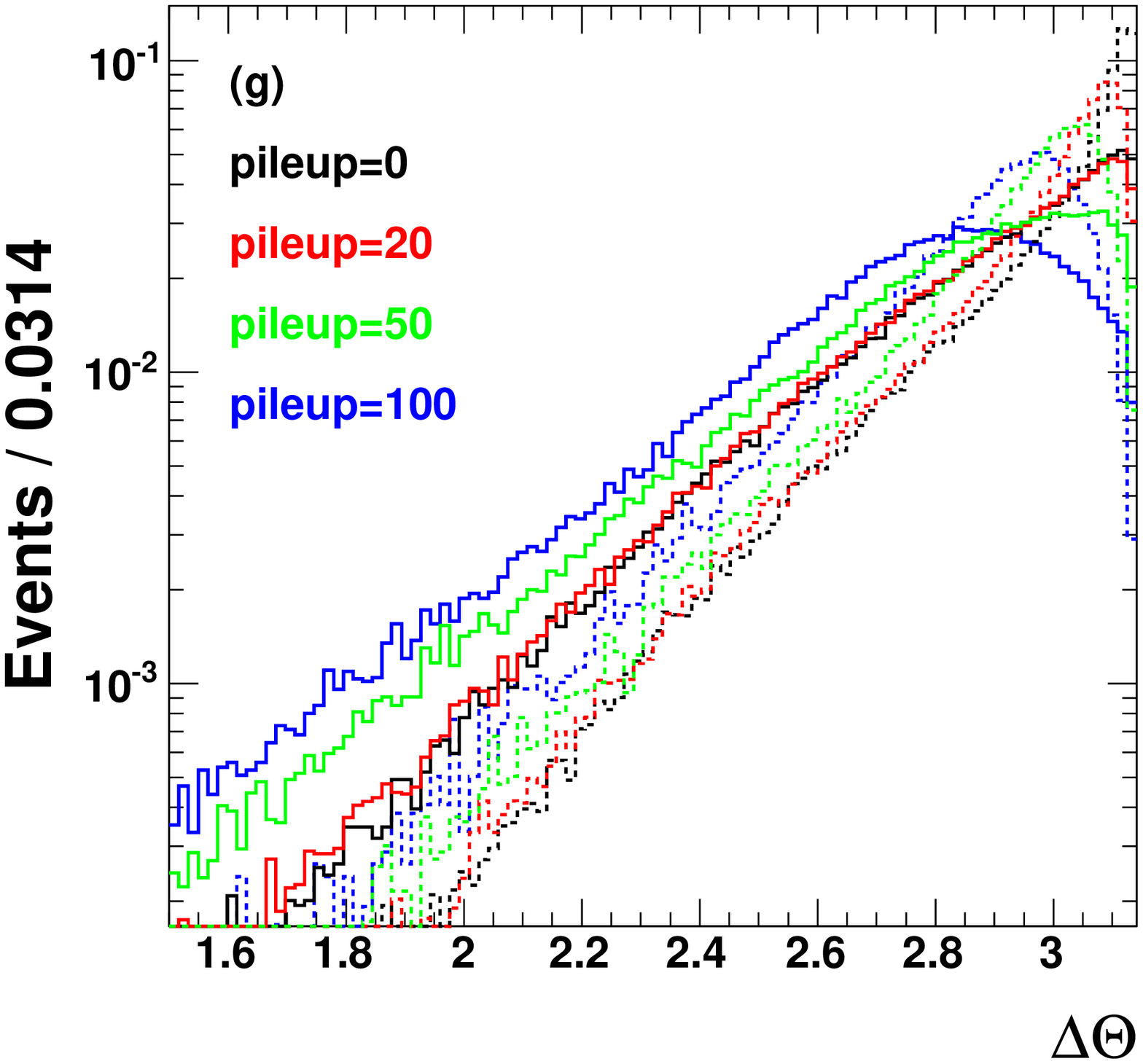}
\includegraphics[width=0.24\textwidth]{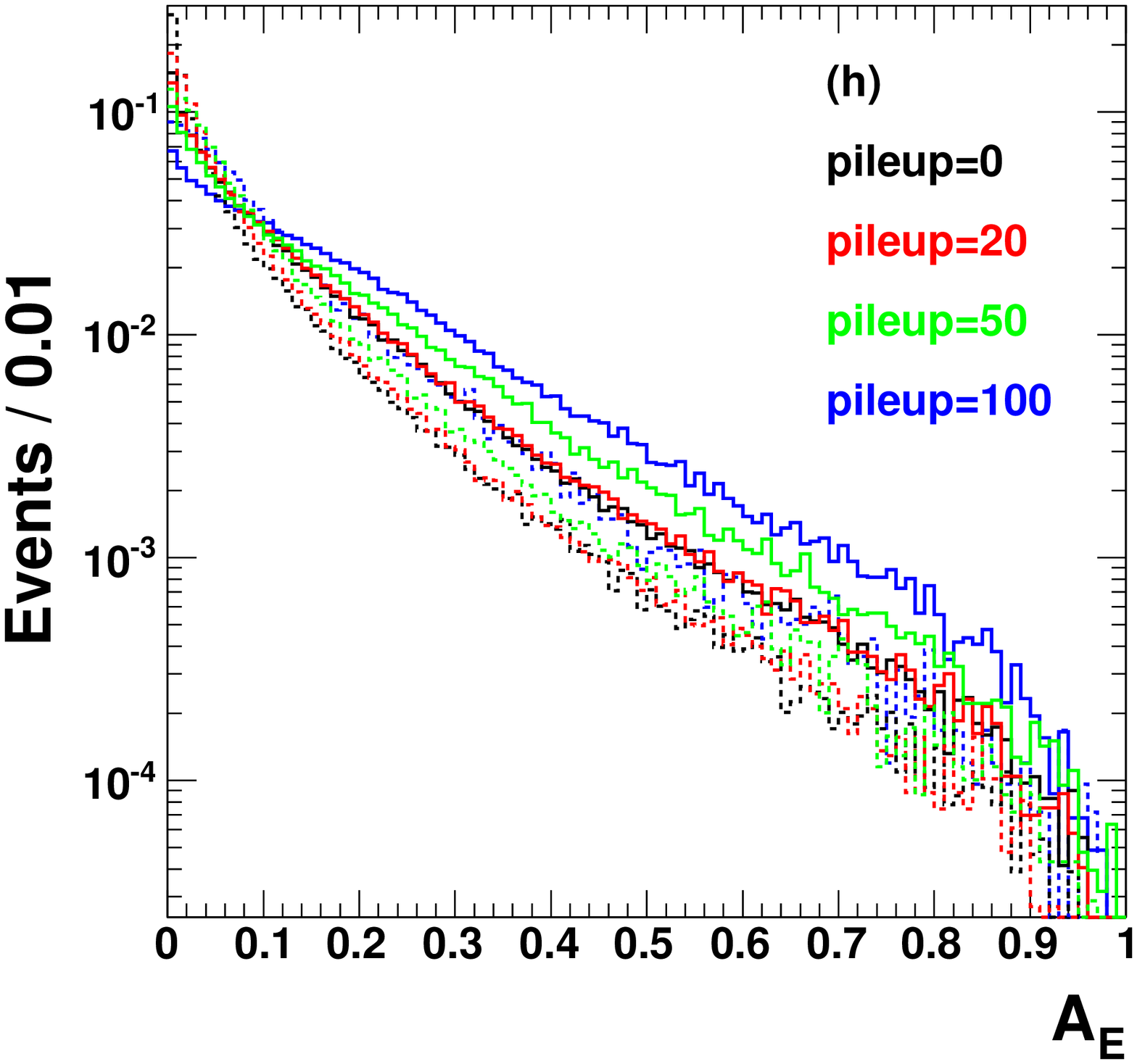}
\caption{The distributions of the jet substructure variables defined in the text for the MC simulated 
QCD jet background (solid line) and $t$ jet signal (dashed line) from the SM $t\bar{t}$ production under different pileup conditions.
The jets used in the evaluation of the jet substructure performance are 
required to have $p_{\rm T}\ge600\,\gev$, $50\,\gev\le m_{\rm jet}\le 350\,\gev$
and at least 3 subjets with $E_{\rm jet}>10\,\gev$ in its rest frame.
All the distributions are normalized to
unity.}
\label{fig:shape}
\vspace{1cm}
\includegraphics[width=0.24\textwidth]{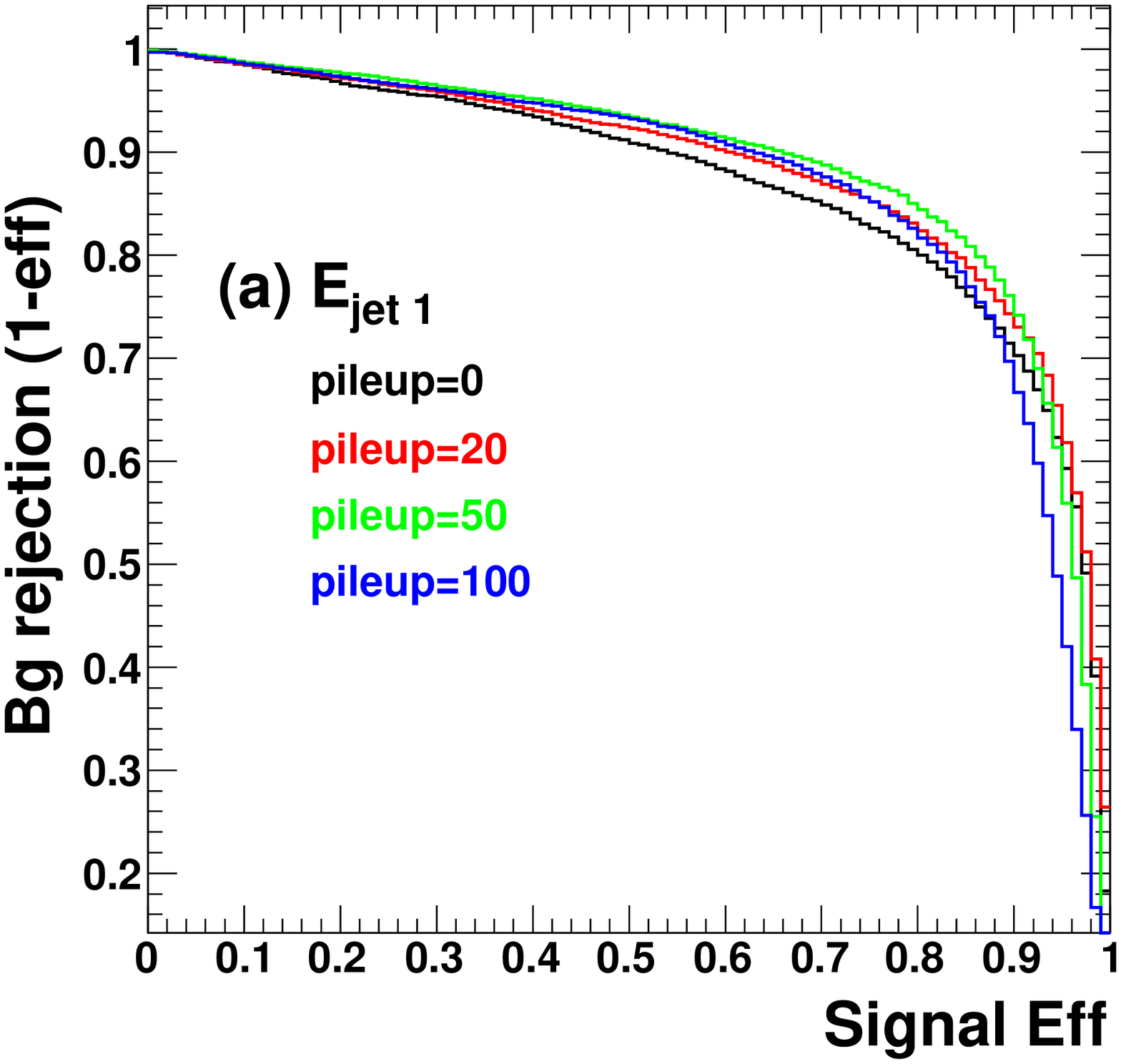}
\includegraphics[width=0.24\textwidth]{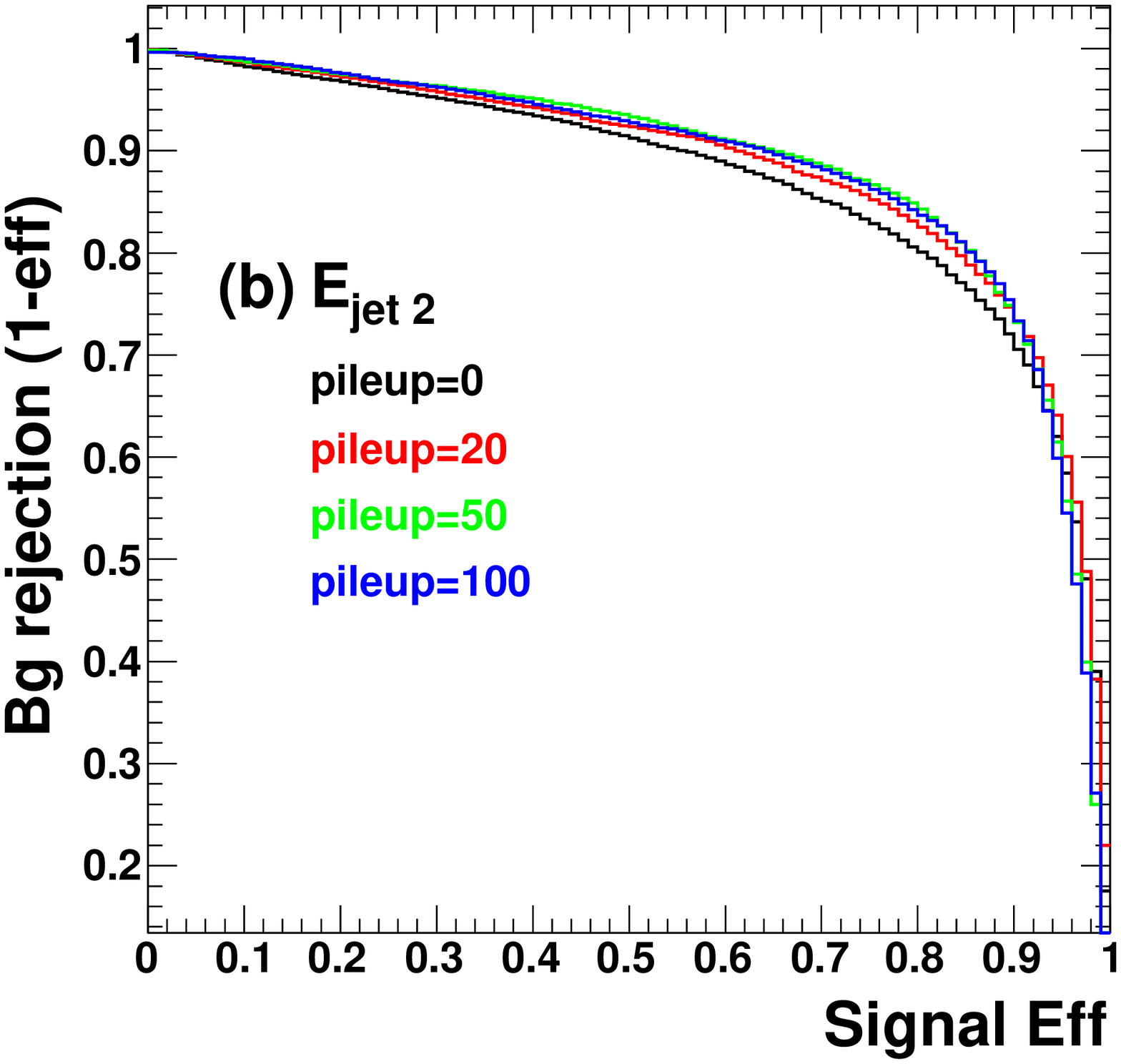}
\includegraphics[width=0.24\textwidth]{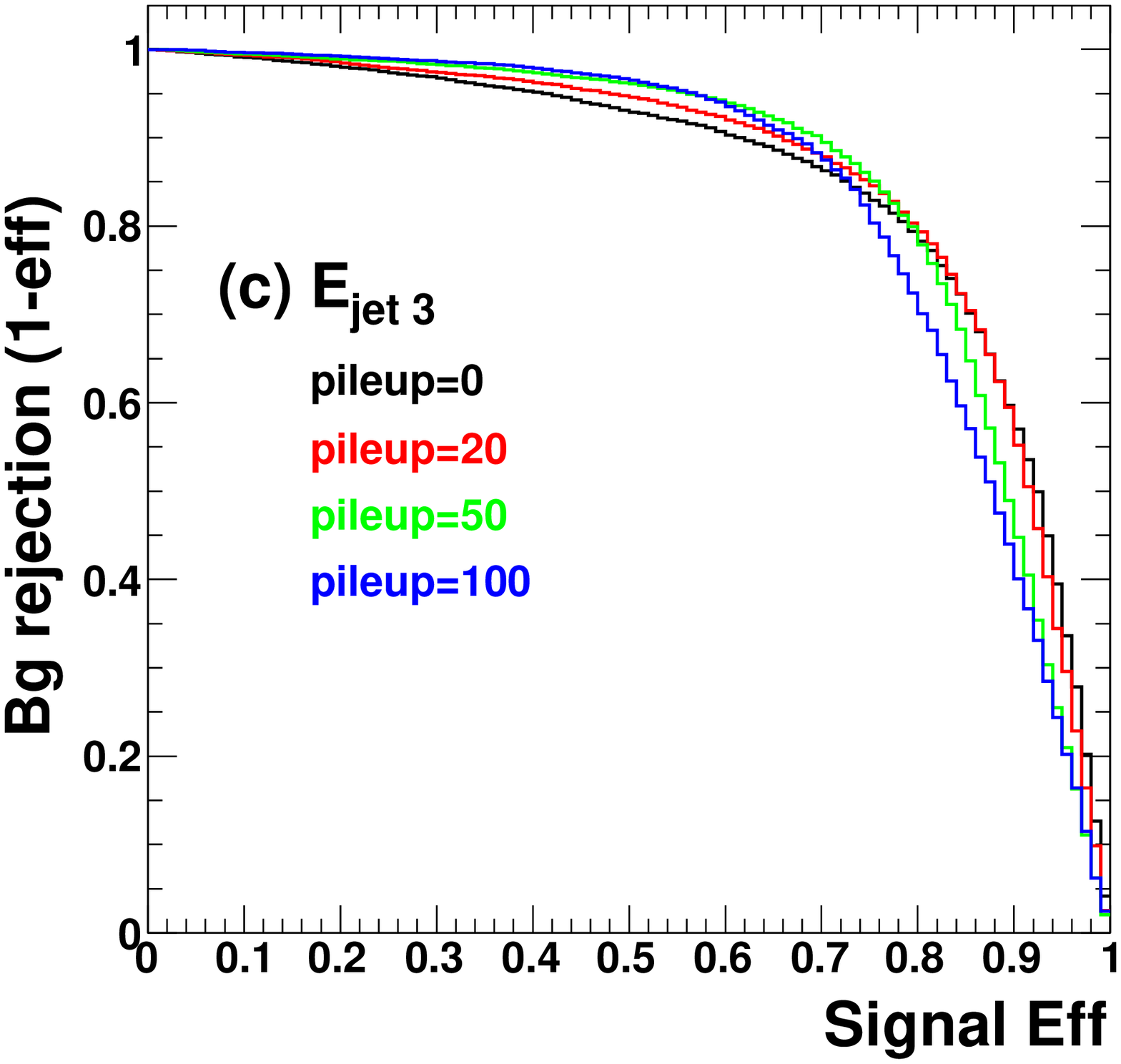}
\includegraphics[width=0.24\textwidth]{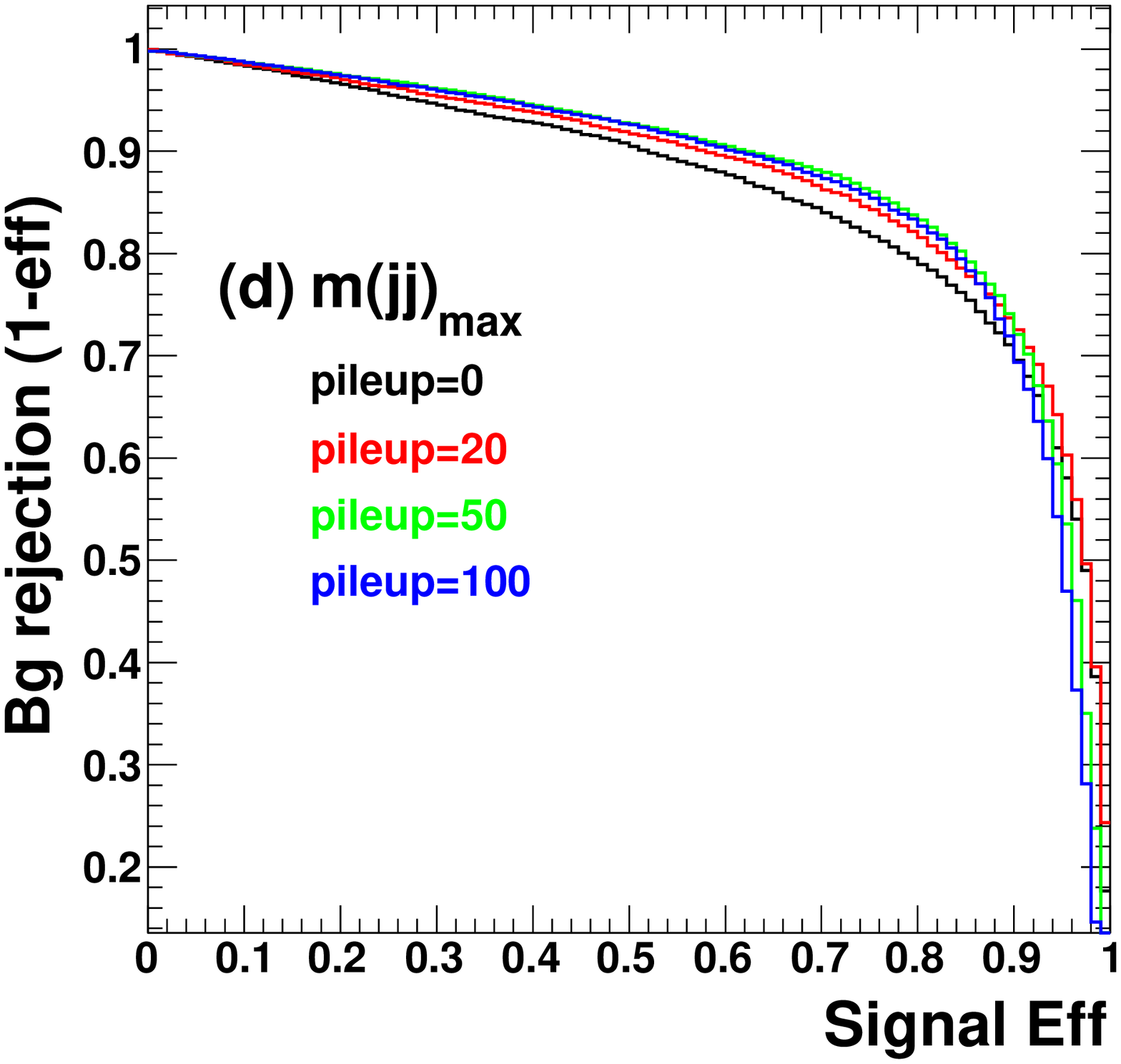}
\includegraphics[width=0.24\textwidth]{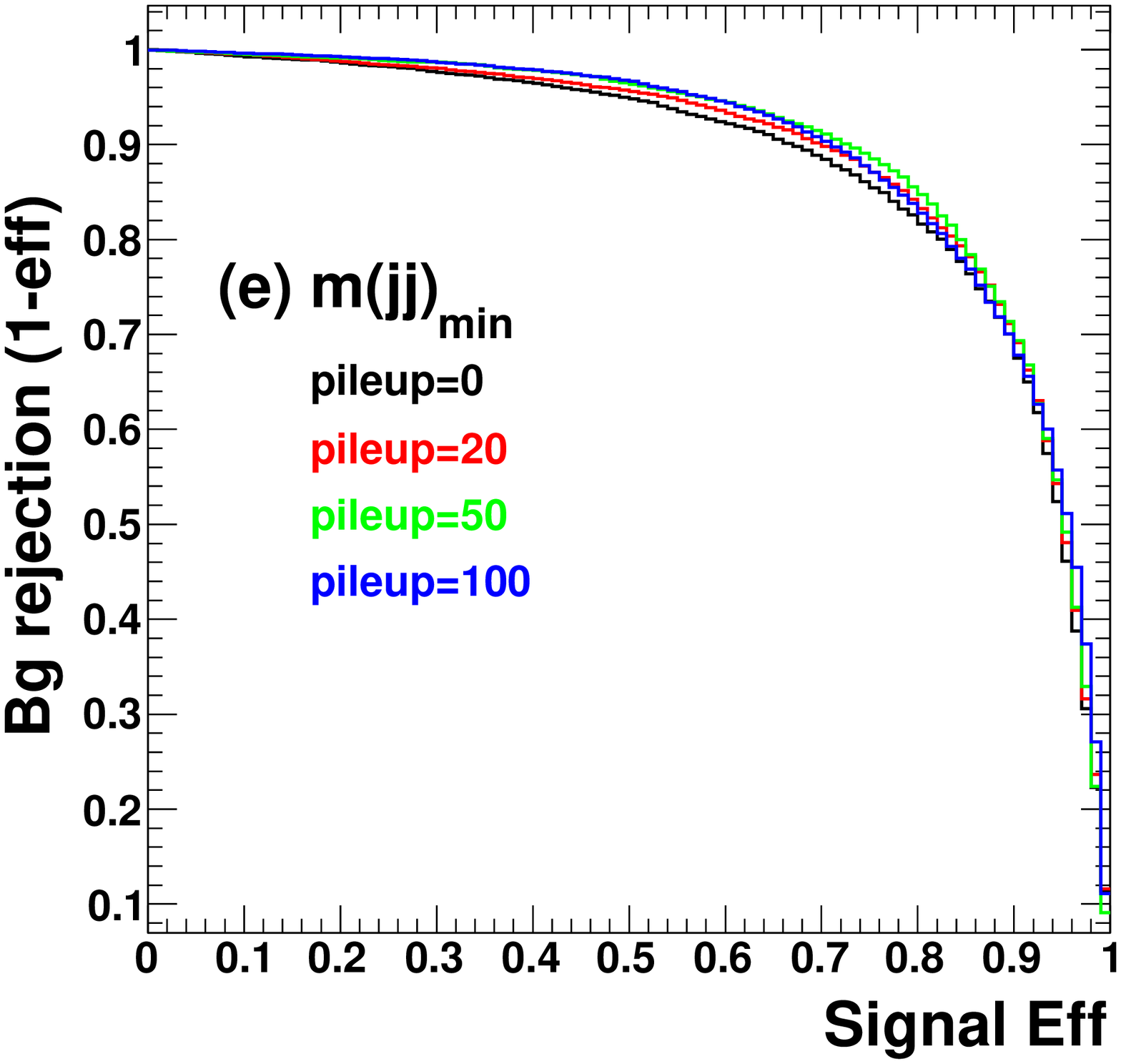}
\includegraphics[width=0.24\textwidth]{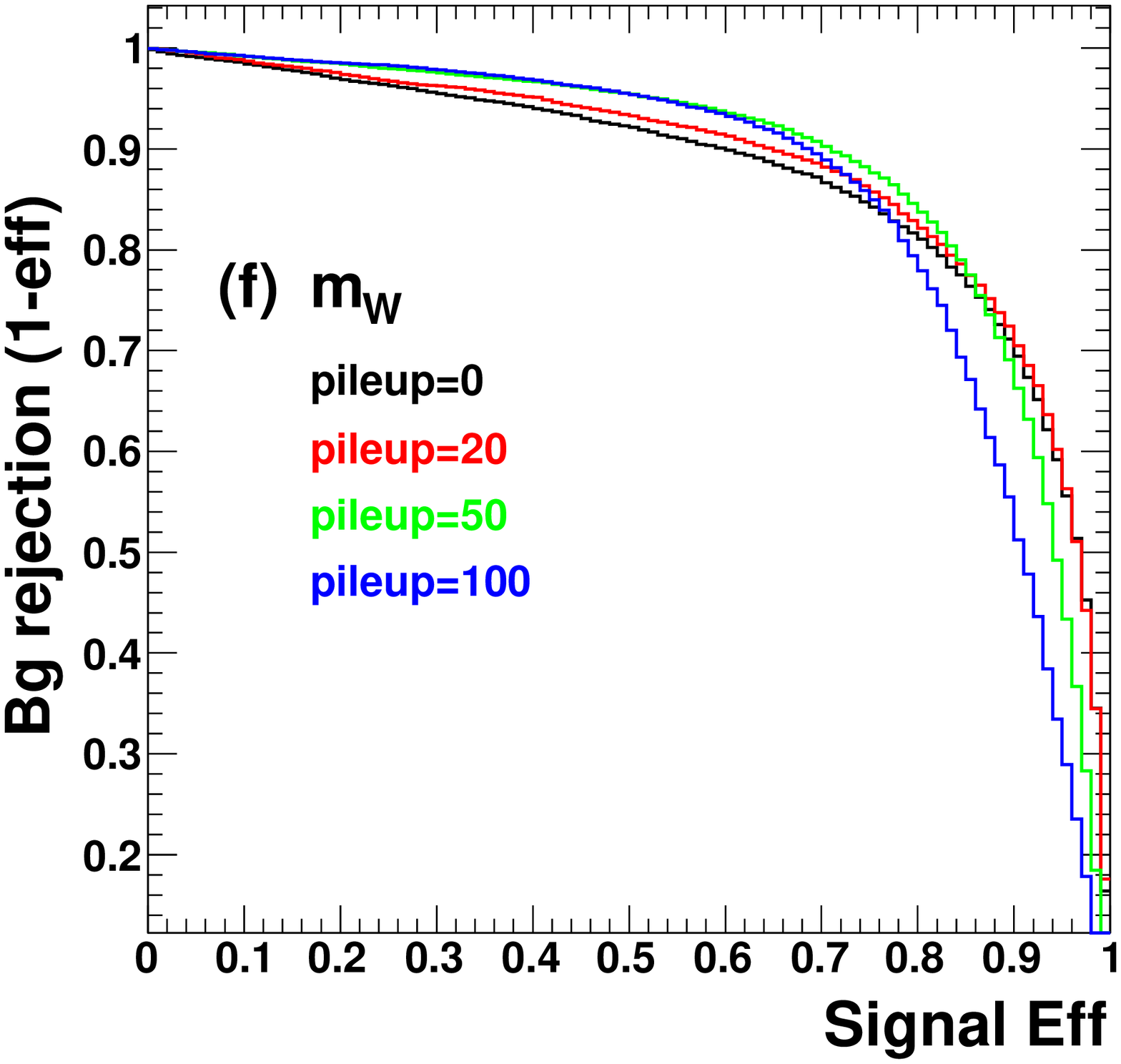}
\includegraphics[width=0.24\textwidth]{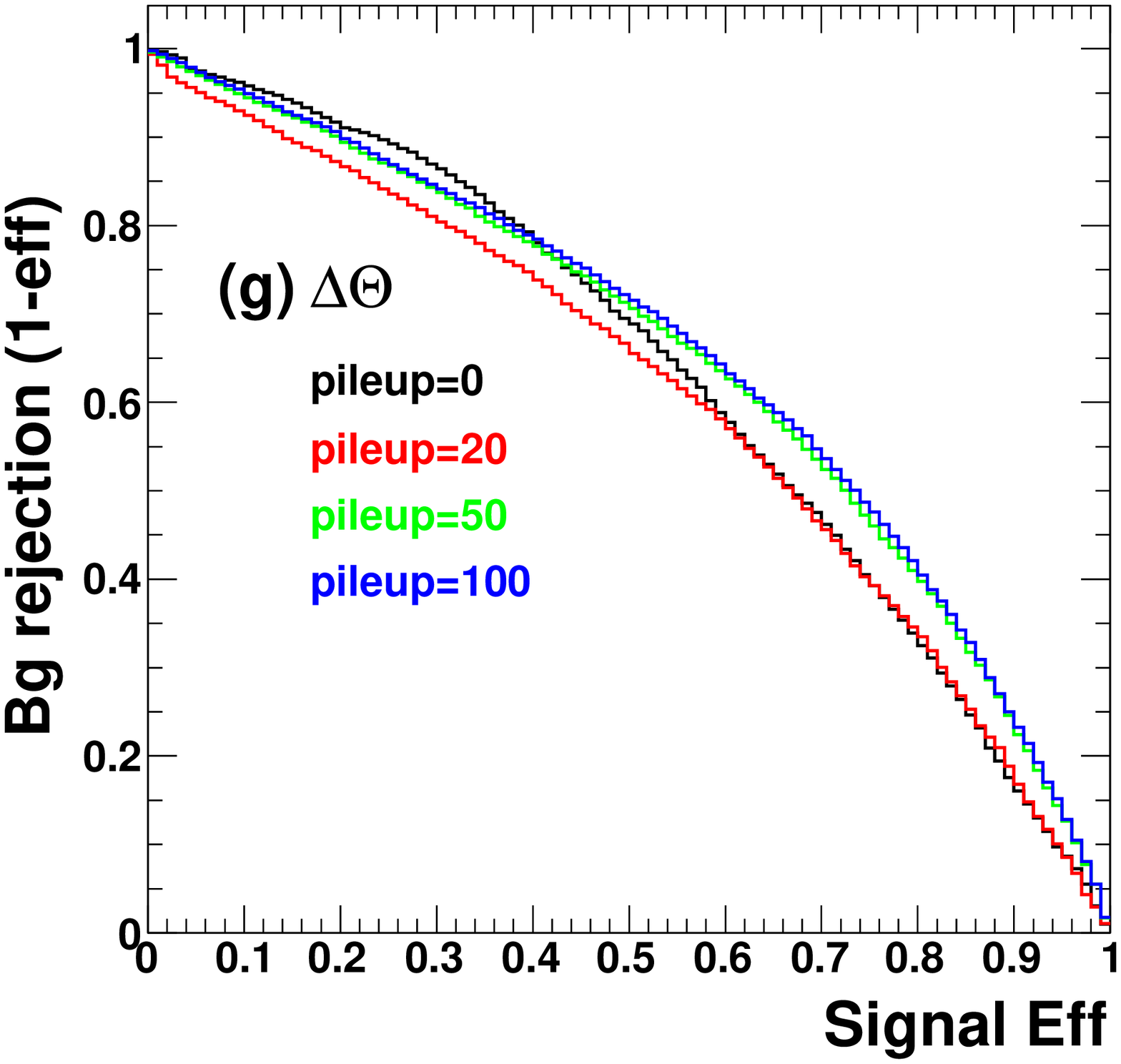}
\includegraphics[width=0.24\textwidth]{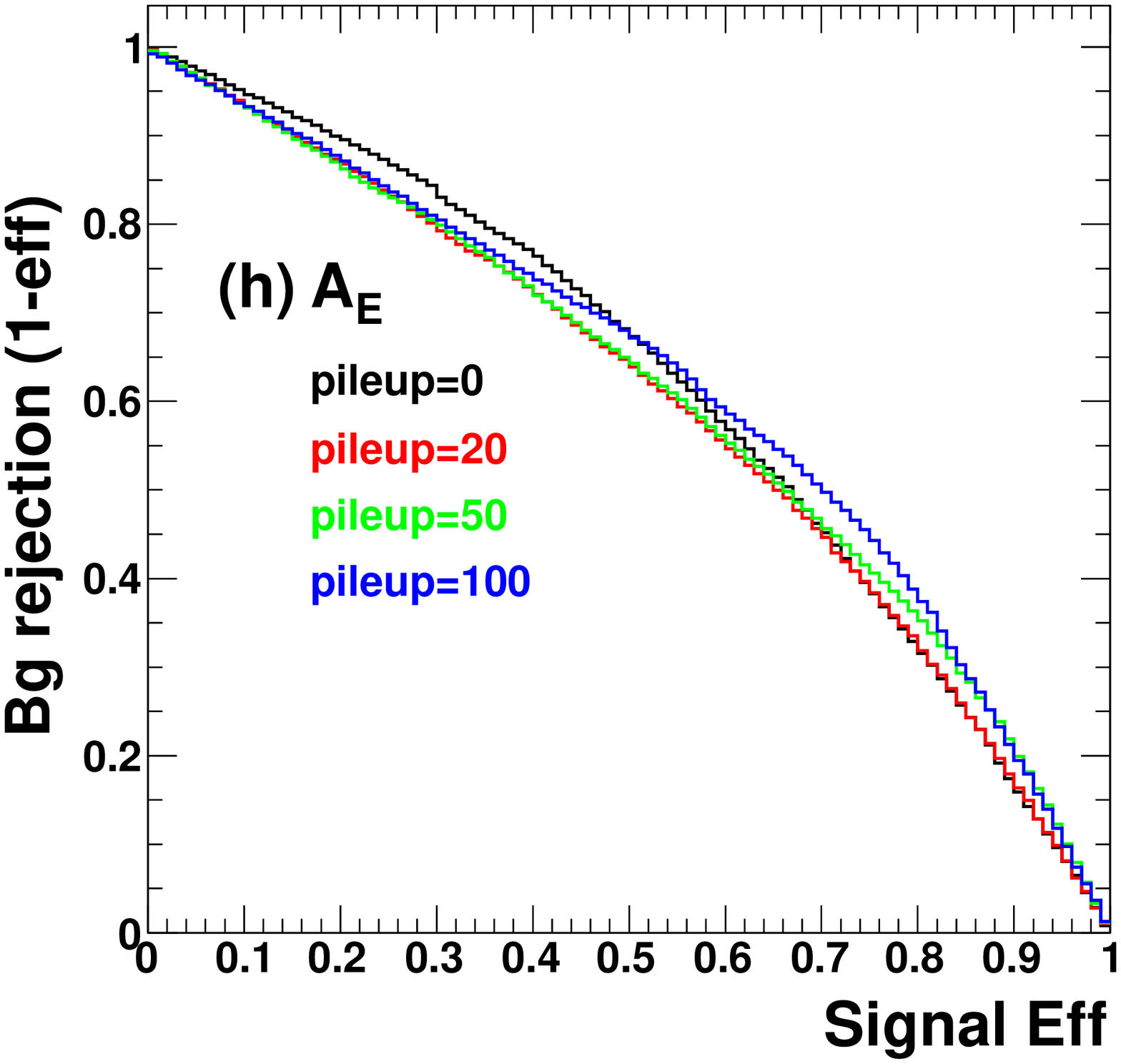}
\caption{The background rejection of QCD jets vs. the signal efficiency of $t$ jets for the jet substructure variables in
different pileup conditions. The jets used in the evaluation of the jet substructure performance are 
required to have $p_{\rm T}\ge600\,\gev$, $50\,\gev\le m_{\rm jet}\le 350\,\gev$
and at least 3 subjets with $E_{\rm jet}>10\,\gev$ in its rest frame.}
\label{fig:eff}
\end{center}
\end{figure*}

\subsection{Jet substructure variables}
We introduce several jet substructure variables to identify boosted $t$ jets.
All of them are calculated using the subjet information  in the jet rest frame. They are:
\begin{itemize}
\item $E_{\rm j_1}$, $E_{\rm j_2}$ and $E_{\rm j_3}$: $E_{\rm j_i}$ is the energy 
of the subject $j_i$ in the center-of-mass frame of the jet. The $j_1$, $j_2$ and $j_3$
denote  the subjets with the first, second and third highest energy.
\item $m_W$: it is defined as $m_{j_1,j_3}$, the invariant mass of the combination of subjet $j_1$ and $j_3$ in the jet rest frame.
\item $m(jj)_{\rm max}$ and $m(jj)_{\rm min}$: they are defined as 
$m(jj)_{\rm max}\equiv \max(m_{j_1,j_2},m_{j_2,j_3})$ and
$m(jj)_{\rm min}\equiv \min(m_{j_1,j_2},m_{j_2,j_3})$.
\item Asymmetry of the energy $A_E$: it is defined as
$A_E=(E_W - E_{\rm j_2})/(E_W + E_{\rm j_2})$, where $E_W$ is the sum of the energies $E_{\rm j_1}$ and $E_{\rm j_3}$.
\item $\Delta\Theta$: it is defined as the opening angle between 
$\vec{p}_{j_1}+\vec{p}_{j_3}$ and $\vec{p}_{j_2}$, where $\vec{p}_{j_i},\;i=1, 2$ and  $3$,
is the momentum of the subjet $j_i$ in the jet rest frame.
\end{itemize}
The distributions of the jet substructure variables are shown
in Fig.~\ref{fig:shape} for $t$ jet signal and QCD jet background
under different pileup conditions.
The corresponding signal efficiencies of $t$ jets vs. the background rejections of QCD jets are 
shown in Fig.~\ref{fig:eff}.
We see a significant difference between the signal and background
distributions for those jet substructure variables. All of them show some dependence on the
pileup conditions.  The effects are slightly less for the $t$ jet
signal than the QCD jet background.  As shown in Fig.~\ref{fig:eff}, although the
performance of the jet substructure variables varies with respect to different 
average number of multiple interaction per event, it does not show any significant
degradation of its rejection power of the QCD jet background while retaining the same
$t$ jet signal identification efficiency.

\subsection{Boosted decision tree top tagger}
\label{sec:BDT}
\begin{figure}[!htb]
\begin{center}
\includegraphics[width=0.4\textwidth]{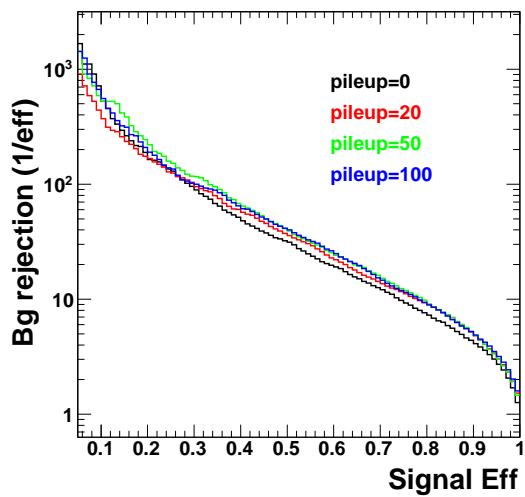}
\caption{The background rejection of QCD jets vs. the signal efficiency of $t$ jets 
 for the boosted decision tree top tagger in
different pileup conditions.
The jets used in the evaluation of the tagger performance are 
required to have $p_{\rm T}\ge600\,\gev$, $50\,\gev\le m_{\rm jet}\le 350\,\gev$
and at least 3 subjets with $E_{\rm jet}>10\,\gev$ in its rest frame.
}
\label{fig:eff_BDT}
\end{center}
\end{figure}
Most jet substructure variables that we introduced in previous section are strongly correlated with each 
other. In order to unitize the maximum discriminating powers of those variables, multivariable 
analysis techniques, such as neural network, boosted decision tree (BDT), etc, are typically
necessary to combine their informations. In this section,
we demonstrate such an application by constructing a BDT variable using the following
jet substructure quantities: $E_{\rm j_1}$, $E_{\rm j_2}$, 
$E_{\rm j_3}$, $m_W$, $m(jj)_{\rm max}$, $m(jj)_{\rm min}$, $A_E$ and $\Delta\Theta$. 
Note that the signal and background separation power of the variables $A_E$ and $\Delta\Theta$
is not as good as the others, as shown in Fig.~\ref{fig:shape} and~\ref{fig:eff}. However, 
their correlations with the other variables are relatively small (less than 30\,\% in most cases). 
As a result, we keep them in the BDT algorithm to add additional discriminating power.
We study the BDT algorithm using MC events generated with different average number
of multiple interactions per event and optimize their performances separately.
The signal efficiency of $t$ jets vs. the background rejection of QCD jets for the BDT variable 
is shown in Fig.~\ref{fig:eff_BDT}. Regardless of the pileup conditions, the jet substructure
variables we proposed can easily reduce the contribution of the QCD jet background by 
approximately 100, with only a factor of three reduction for the $t$ jet signal identification efficiency.
In principle, we expect that the tagger performance gets worse when the pileup increases.
As shown in Fig.~\ref{fig:eff_BDT}, the performance of the tagger is actually better with higher
pileups for several ranges of the signal efficiencies. This is an artificial effect that is caused by the
selection of jets used in the evaluation of the tagger performance. In our studies, we 
only use jets that have $p_{\rm T}\ge600\,\gev$, $50\,\gev\le m_{\rm jet}\le 350\,\gev$
and at least 3 subjets with $E_{\rm jet}>10\,\gev$ in its rest frame. As a result, when pileup
increases, many QCD jets that otherwise would not satisfy  the jet selection criteria are selected.

Comparing to the performance of other existing top taggers~\cite{Altheimer:2012mn},
the BDT top tagger based on jet substructure in the jet rest frame has similar background rejection
for a given signal identification efficiency of boosted hadronically decaying top quarks. 
In our studies, the jets are reconstructed using a distance parameter $\Delta R=0.6$ that is 
different from many other studies. We also consider the pileup conditions in extreme cases,
which is absence in many of performance studies of existing top taggers. Our results show 
that top tagger based on jet substructure in the jet rest frame is complementary to
other top tagger tools.

\section{Application}
\label{sec:app}
Reconstructed decays of top quarks to jets can be used
to search for NP with specific final state signatures. Here
we demonstrate such an application by considering a heavy
resonance that decays to a $t\bar{t}$ final state: $pp\to X\to t\bar{t}$, 
where $X$ is a new heavy resonance beyond the SM, such as a new heavy gauge boson $Z^\prime$ 
or a KK gluon in the Randall-Sundrum model, etc. 

We consider a search for an $X$ signal by fully reconstructing
the $X$ signal candidate in the decay mode where
both top quarks decay hadronically. Note that for such high mass ($>2.0\,\tev$)
resonance decays, more than 90\,\% of the events have 
their decay particles from the top decay within a cone of $R<0.6$. This makes it very difficult to identify separate subjets
from top quark decays in the lab frame. As a result, we select the two leading jets with 
the highest and second highest energy with $p_{\rm T}>600\,\gev$ and $|\eta|<1.9$ in an event as the two
hadronically decaying top quark candidates. The jets are reconstructed using the anti-$k_{\rm T}$ algorithm with a distance parameter 
$\Delta R=0.6$. We then subsequently combine the two top jet candidates to form the candidate of the $X\to t\bar{t}$ 
resonance.
\begin{figure}[!htb]
\begin{center}
\includegraphics[width=0.4\textwidth]{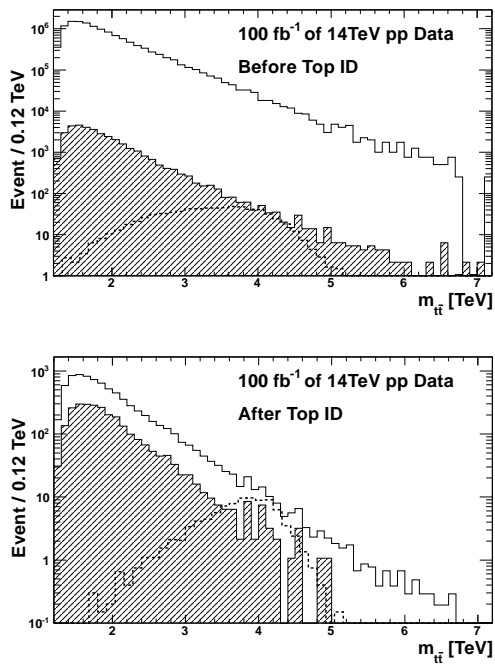}
\caption{The invariant mass distribution of the $X\to t\bar{t}$ candidates in the MC simulated event sample that is
equivalent to $100\,\ifb$ of LHC data at $14\,\tev$ center-of-mass energy before (top) and
after (bottom) the BDT top tagger is applied. The MC events used to make the plot are generated with an assumption 
of 50 multiple interactions per event on average. The open histogram is the background distribution from
QCD events, the hatched histogram is the background contribution from the SM $t\bar{t}$ production, and the dashed histogram
is the expected signal distribution from $pp\to X\to t\bar{t}$. Here we assume a width of 100\,\gev\ of the heavy resonance with $m_X=4\,\tev$, 
and the product of the production cross section of $X$
 and the branching fraction for its decay into $t\bar{t}$ pair is 10 fb.}
\label{fig:mGtt}
\end{center}
\end{figure}

The invariant mass distributions of the $X\to t\bar{t}$ candidates in the MC simulated event sample that is equivalent to
$100\,\ifb$ of LHC data at 14\,\tev\ center-of-mass energy are shown in Fig.~\ref{fig:mGtt}. Here we consider that the
average pileup is 50 and the mass of the heavy resonance is $m_X=4\,\tev$ with a natural width of $100\,\gev$.  The major SM backgrounds
are QCD events and $t\bar{t}$ production. In order to reduce the background, we apply the hadronic top jet identification
using the BDT top tagger described in Section~\ref{sec:BDT}. The selection criterion of the BDT variable is optimized to have 
approximately $40\,\%$ identification efficiency of a top jet, while keeping the fake rate of QCD jets at less than 2\,\%.
As the result, we are able to reduce the SM background by a factor of more than 2000, while keeping roughly $15\,\%$ of
the signal events. As shown in Fig.~\ref{fig:mGtt}, such a selection dramatically improves the signal over background
ratio and makes it possible for us to observe a potential $X\to t\bar{t}$ signal if the
product of its production cross section  and the branching fraction for its decay into a $t\bar{t}$ pair is of the order of a few tens of $\rm fb$.

We repeat the study for heavy resonance $X$ with different masses under various pileup conditions.
We estimate the expected 95\,\% C.L. upper limit on the product of the production cross section
of a heavy resonance $X$ and the branching faction for its decay into a $t\bar{t}$ pair. The expected limit for
100\,\ifb\ of LHC data at 14\,\tev\ center-of-mass energy is plotted as a function of the assumed $X$ mass, as 
shown in Fig.~\ref{fig:limit}.   All the studies show similar results. The BDT top tagger based on the jet substructure information
in the jet rest frame can significantly improve  our experimental 
sensitivities in search for $pp\to X\to t\bar{t}$. The performance of the top tagger decreases when the average number of
multiple interactions per event increases. However, the degradation is rather small. Even in the case of an extreme
pileup condition with 100 multiple interactions per event, the experimental sensitivity drops by less than a factor of 2 compared to the
one in an ideal experiment with zero pileup.
\begin{figure}[!htb]
\begin{center}
\includegraphics[width=0.42\textwidth]{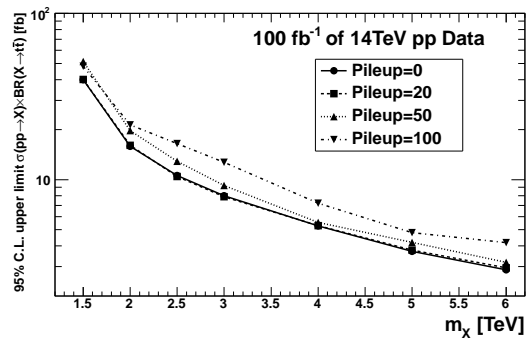}
\caption{The expected 95\,\% C.L. upper limit on the product of the production cross section
of a heavy resonance $X$ and its decaying branching faction into a $t\bar{t}$ pair, as  a function of the assumed $X$ mass
under different pileup conditions.
Here we assume a width of 100\,\gev\ of the heavy resonance and 100\,\ifb\ of LHC data at
14\,\tev\ center-of-mass energy.}
\label{fig:limit}
\end{center}
\end{figure}

\section{Conclusion}
\label{sec:conclusion}
In this paper we study the identification of a boosted hadronically decaying top quark
using jet substructure in the center-of-mass frame of the jet. 
We demonstrate that the method can greatly reduce the QCD jet background while maintaining high
identification efficiency of the boosted hadronically decaying top quarks even in a very high
pileup condition. 
The study shows a  good prospective on search for heavy mass particles in the $t\bar{t}$ decay channel 
for the LHC experiments at 14\,\tev\ center-of-mass energy.

\section{Acknowledgments}
We thank Jim Cochran, Nils Krumnack, Soeren Prell and German Valencia 
for many discussions and valuable comments on the manuscript. 
During the review process of this paper, it has been pointed out 
that similar ideas were also studied by Gavin Salam 
but they were not published. This work is
supported by
the Office of Science of the U.S. Department of Energy
under contract DE-FG02-12ER41827 and by US ATLAS collaboration.


\begin{thebibliography}{99}

\bibitem{Agashe:2006hk} 
  K.~Agashe, A.~Belyaev, T.~Krupovnickas, G.~Perez and J.~Virzi,
  Phys.\ Rev.\ D {\bf 77}, 015003 (2008).
  
\bibitem{Contino:2006nn} 
  R.~Contino, T.~Kramer, M.~Son and R.~Sundrum,
  JHEP {\bf 0705}, 074 (2007).
    
\bibitem{Matsumoto:2006ws} 
  S.~Matsumoto, M.~M.~Nojiri and D.~Nomura,
  Phys.\ Rev.\ D {\bf 75}, 055006 (2007).
  
\bibitem{Fitzpatrick:2007qr} 
  A.~L.~Fitzpatrick, J.~Kaplan, L.~Randall and L.~-T.~Wang,
  JHEP {\bf 0709}, 013 (2007).
      
\bibitem{Thaler:2008ju}
  J.~Thaler and L.~T.~Wang,
  J. High Energy Phys. {\bf 0807}, 092 (2008).

\bibitem{Kaplan:2008ie}
  D.~E.~Kaplan, K.~Rehermann, M.~D.~Schwartz and B.~Tweedie,
  Phys.\ Rev.\ Lett.\  {\bf 101}, 142001 (2008).

\bibitem{Almeida:2008tp}
  L.~G.~Almeida, S.~J.~Lee, G.~Perez, I.~Sung and J.~Virzi,
  Phys.\ Rev.\  D {\bf 79}, 074012 (2009).

\bibitem{Krohn:2009wm}
  D.~Krohn, J.~Shelton and L.~T.~Wang,
  J. High Energy Phys. {\bf 1007}, 041 (2010).

\bibitem{Chekanov:2010vc}
  S.~Chekanov and J.~Proudfoot,
  Phys.\ Rev.\  D {\bf 81}, 114038 (2010).

\bibitem{Plehn:2010st}
  T.~Plehn, M.~Spannowsky, M.~Takeuchi, D.~Zerwas,
  J. High Energy Phys. {\bf 1010}, 078 (2010).

\bibitem{Bhattacherjee:2010za}
  B.~Bhattacherjee, M.~Guchait, S.~Raychaudhuri and K.~Sridhar,
  Phys.\ Rev.\  D {\bf 82}, 055006 (2010).

\bibitem{Rehermann:2010vq}
  K.~Rehermann, B.~Tweedie,
  J. High Energy Phys. {\bf 1103}, 059 (2011).

\bibitem{Chekanov:2010gv}
  S.~Chekanov, C.~Levy, J.~Proudfoot and R.~Yoshida,
  Phys.\ Rev.\  D {\bf 82}, 094029 (2010)

\bibitem{Jankowiak:2011qa}
M.~Jankowiak and A.~J.~Larkoski,
JHEP {\bf 1106}, 057 (2011).

\bibitem{Thaler:2011gf}
 J.~Thaler and K.~Van Tilburg,
JHEP {\bf 1202}, 093 (2012).

\bibitem{Soper:2012pb}
D.~E.~Soper and M.~Spannowsky,
arXiv:1211.3140 [hep-ph].

\bibitem{Aad:2012raa} 
[ATLAS Collaboration],
  JHEP {\bf 1301}, 116 (2013).
  
  
\bibitem{Chatrchyan:2012ku} 
[CMS Collaboration],
  JHEP {\bf 1209}, 029 (2012).  
  
\bibitem{Abdesselam:2010pt}
 A.~Abdesselam, E.~B.~Kuutmann, U.~Bitenc, G.~Brooijmans, J.~Butterworth, P.~Bruckman de Renstrom, D.~Buarque Franzosi, R.~Buckingham {\it et al.},
  Eur.\ Phys.\ J.\  {\bf C71}, 1661 (2011).
    
\bibitem{Butterworth:2008iy} 
  J.~M.~Butterworth, A.~R.~Davison, M.~Rubin and G.~P.~Salam,
  Phys.\ Rev.\ Lett.\  {\bf 100}, 242001 (2008).
   
\bibitem{Ellis:2009su} 
  S.~D.~Ellis, C.~K.~Vermilion and J.~R.~Walsh,
  Phys.\ Rev.\ D {\bf 80}, 051501 (2009).

\bibitem{Ellis:2009me} 
  S.~D.~Ellis, C.~K.~Vermilion and J.~R.~Walsh,
  Phys.\ Rev.\ D {\bf 81}, 094023 (2010).
      
 \bibitem{Krohn:2009th}
  D.~Krohn, J.~Thaler, L.~-T.~Wang,
  J. High Energy Phys. {\bf 1002}, 084 (2010).
  
\bibitem{Chen:2011ah} 
  C.~Chen,
  Phys.\ Rev.\ D {\bf 85}, 034007 (2012).

\bibitem{Kim:2010uj}
  J.~-H.~Kim,
  Phys.\ Rev.\  {\bf D83}, 011502 (2011).

\bibitem{Sjostrand:2006za}
  T.~Sj{\"o}strand, S.~Mrenna, P.~Z.~Skands,
 J. High Energy Phys. {\bf 0605}, 026 (2006).

\bibitem{fastjet}
M.~Cacciari and G.~P.~Salam,
  Phys. Lett. {\bf 641}, 57 (2006).
  
\bibitem{Cacciari:2008gp}
M.~Cacciari, G.~P.~Salam and G.~Soyez,
  J. High Energy Phys. {\bf 0804}, 063 (2008).
 
 \bibitem{Dokshitzer:1997in} 
  Y.~L.~Dokshitzer, G.~D.~Leder, S.~Moretti and B.~R.~Webber,
  JHEP {\bf 9708}, 001 (1997).
  
 \bibitem{Altheimer:2012mn} 
  A.~Altheimer, S.~Arora, L.~Asquith, G.~Brooijmans, J.~Butterworth, M.~Campanelli, B.~Chapleau and A.~E.~Cholakian {\it et al.},
  J.\ Phys.\ G {\bf 39}, 063001 (2012).
   
\end{thebibliography}
\end{document}